\documentclass[fleqn,usenatbib]{mnras}

\usepackage{newtxtext,newtxmath}
\usepackage{graphics,graphicx}
\usepackage{epstopdf}
\usepackage{latexsym}
\usepackage{amssymb}
\usepackage{amsmath}
\usepackage{url}

\usepackage{graphicx}
\usepackage{times}
\usepackage{natbib}

\begin{document}
 
\title[Discovery of large-scale ultra-steep diffuse radio emission in low-mass cluster Abell 1931]{Discovery of large-scale diffuse radio emission in low-mass galaxy cluster Abell 1931}

\author[M. Br\"{u}ggen et al.]{M. Br\"{u}ggen$^{1}$\thanks{E-mail: mbrueggen@hs.uni-hamburg.de}, 
D. Rafferty$^{1}$, 
A. Bonafede$^{1,2}$, 
R. J. van Weeren$^{3,4}$, 
T. Shimwell$^{3}$, 
\newauthor
H. Intema$^{3}$,
H. R\"ottgering$^{3}$,
G. Brunetti$^{2}$,
G. Di Gennaro$^{4,3}$, 
F. Savini$^{1}$,
A. Wilber$^{1}$
\newauthor
S. O'Sullivan$^{1}$,
T. A. Ensslin$^{5}$,
F. De Gasperin$^{3,1}$,
M. Hoeft$^{6}$
\\
$^{1}$ Hamburger Sternwarte, Gojenbergsweg 112, 21029 Hamburg, Germany\\
$^{2}$ Istituto di Radio Astronomia, INAF, Via Godetti 101, 40121 Bologna, Italy\\
$^{3}$ Leiden Observatory, Leiden University, P.O. Box 9513, NL-2300 RA Leiden, The Netherlands\\
$^{4}$ Harvard-Smithsonian Center for Astrophysics, 60 Garden Street, Cambridge, MA 02138, USA\\
$^{5}$ Max-Planck-Institut f\"{u}r Astrophysik,
Karl-Schwarzschild-Str.1, Postfach 1317, 85741 Garching, Germany\\
$^{6}$ Th{\"u}ringer Landessternwarte, Sternwarte 5, 07778 Tautenburg, Germany
}

\date{Accepted ???. Received ???; in original form ???}
\maketitle

\begin{abstract}
Extended, steep-spectrum radio synchrotron sources are pre-dominantly found in massive galaxy clusters as opposed to groups.
LOFAR Two-Metre Sky Survey images have revealed a diffuse, ultra-steep spectrum radio source in the low-mass cluster Abell 1931. The source has a fairly irregular morphology with a largest linear size of about 550 kpc. The source is only seen in LOFAR observations at 143 MHz and GMRT observations at 325 MHz. The spectral index of the total source between 143 MHz and 325 MHz is $\alpha_{143}^{325} = -2.86 \pm 0.36$. The source remains invisible in Very Large Array (1--2 GHz) observations as expected given the spectral index. Chandra X-ray observations of the cluster revealed a bolometric luminosity of $L_X = (1.65 \pm 0.39) \times 10^{43}$ erg s$^{-1}$ and a temperature of $2.92_{-0.87}^{+1.89}$ keV which implies a mass of around $\sim 10^{14} M_{\odot}$. We conclude that the source is a remnant radio galaxy that has shut off around 200 Myr ago. The brightest cluster galaxy, a radio-loud elliptical galaxy, could be the source for this extinct source. Unlike remnant sources studied in the literature, our source has a steep spectrum at low radio frequencies. Studying such remnant radio galaxies at low radio frequencies is important for understanding the scarcity of such sources and their role in feedback processes.
\end{abstract}
\label{firstpage}

\section{Introduction}
\label{sec:intro}

Diffuse radio sources in galaxy clusters, radio relics and halos, are signatures of dynamic cluster events. Radio relics are low surface brightness ($\simeq~10^{-6}$ Jy/arcsec$^2$ at 1.4~GHz), steep-spectrum\footnote{S($\nu$)$\propto 
\nu^{\alpha}$} ($\alpha < -1$) sources that  are typically located in the periphery of galaxy clusters. While it is known that non-thermal electrons are
responsible for the observed synchrotron radiation, the mechanism of particle acceleration as well as the origins of the magnetic fields on these scales are poorly understood \citep{Brunetti2014, 2017CQGra..34w4001V}.

Radio relics trace low-Mach number shocks ($\mathcal{M} \lesssim 3$) generated by cluster-cluster merger events. A long-standing problem is how low-Mach number shocks can accelerate electrons so efficiently to explain the bright radio relics that are observed.  Current theories of diffusive shock acceleration have difficulties in explaining the required efficiency of particle acceleration in the low-Mach number shocks as found in galaxy clusters \citep[e.g.][]{Kang2011, Guo2014}. In particular, it remains unclear what role pre-existing populations of relativistic electrons play. 
It has been suggested that these fossil electrons, with Lorentz factors of $\sim 200$ and life times of several Gyrs, can be efficiently re-accelerated at shocks and are therefore able to create bright radio relics \citep[][sometimes called "gischt" relics]{1998A&A...332..395E, Kempner2004, Kang2015, Pinzke2013}.\\

Radio halos are Mpc-sized, synchrotron sources that fill the central regions of galaxy clusters.  
They arise from the acceleration of particles by turbulence generated during cluster mergers. Hadronic models that involve the generation of secondary particles via proton-proton collisions in the intracluster medium \citep{Dennison1980} are believed to play a lesser role \citep{Brunetti2014}. However, hadronic models could be important for mini-halos \citep{Pfrommer2004}.\\

Most of the information on the statistical connection between cluster mass, dynamics, and radio halos are obtained by surveys of X-ray luminous galaxy clusters that have been carried out with radio telescopes \citep{Venturi2008, Kale2013, Kale2015}.  There is strong evidence that radio halos and relics depend on the dynamical state of the clusters \cite[e.g.][]{Cuciti2015, Wen2013}. 

In the turbulent re-acceleration model at the low-mass end, only very disturbed clusters host diffuse radio sources and the slope of the $P_{\rm 140\,MHz} - L_{\rm{X}}$ relation steepens at lower luminosities \citep{Cassano2010}.
A subclass of radio halos with ultra-steep spectra mostly lie slightly below the scaling relation, potentially filling in the region between the correlation and the upper limits \citep[e.g.][]{Brunetti2008, Wilber2017}.\\

Another class of steep spectrum radio sources that can be found in galaxy clusters are so-called radio phoenices and AGN relics, which are both related to radio galaxies. AGN relics are essentially extinct or dying radio galaxies. Their radio spectra are steep owing to the synchrotron and inverse Compton losses. Fossil radio plasma from past episodes of AGN activity can also be compressed by shock waves. The ensuing increase in the momentum of the relativistic electrons as well as the increase in the magnetic field strength can produce a radio phoenix \citep{2001A&A...366...26E,Ensslin2002}. Again, these sources have steep and curved radio spectra. The distribution of fossil plasma throughout clusters is an important question since old populations of relativistic electrons are required for most acceleration models that explain radio relics and halos. They also bear witness to past AGN activity and constitute a source of non-thermal pressure in the intra-cluster medium (ICM). The study of past radio-loud AGN activity is important for quantifying of AGN feedback, which is key in preventing catastrophic cooling in the centres of galaxy clusters. \cite{Murgia2011} have studied remnant sources in galaxy clusters and suggest that the duration of the fading phase of remnant radio galaxies in clusters is longer than in the field. The reason may be that the expansion of the radio lobes in clusters has to work against a higher ambient pressure.\\

Based on the number density of radio galaxies, one would expect a large number of remnant radio sources. Yet, observations show that they are exceptionally rare \citep{Giovannini1988,Parma2007}.
Recently, \cite{Brienza2016b} found an AGN relic or remnant radio galaxy in LOFAR images at 150 MHz that extends over 700 kpc but is not associated with a galaxy cluster.  Also its spectrum is not steep at low frequencies and only steepens at frequencies $> 1.4$ GHz.
In follow-up work, \cite{Brienza2017} selected from the LOFAR Lockman Hole field 23 candidates for remnant galaxies, all outside of galaxy clusters. Interestingly, many of these remnants do not have very steep spectra.

\cite{Savini2018} have discovered an extended radio galaxy embedded in steep diffuse emission in a galaxy group. The spectral index map is inconclusive as to whether this source is still active 
Detections of remnant radio galaxies are surprisingly rare.
\cite{2009ApJ...698L.163D} have studied a sample of relic galaxies with the GMRT and the VLA. The find that their sources that show no obvious cores and jets have steep spectra that suggest that the AGN activity has stopped 15-100 million years ago. Before that, these sources were up to 1000 times brighter than their current luminosities making them comparable to the brightest AGN in the local universe.

\cite{Godfrey2017} find that fewer than 2\% of FRII radio galaxies with $S_{\rm 74\, MHz} > 1.5$ Jy are ultra-steep spectrum remnants. They define an ultra-steep spectrum as spectra with indices between 74 MHz and 1.4 GHz of $< -1.2$. 
\cite{Godfrey2017} have modelled the evolution of the remnants in order to match the low number of observed sources. They suggest that the remnants must have a rapid luminosity evolution and keep on adiabatically expanding after the core switches off. This was done by comparing mock samples with observations. In \cite{Godfrey2017} this was done for FR II sources and in \cite{Brienza2017} for FRI sources.

Since radio relics have steep radio spectra, they are best discovered at low radio frequencies. However, until now, only shallow low-frequency radio surveys have been carried out \citep[i.e., the 150 MHz TGSS ADR survey, the 325~MHz WENSS survey and the 74~MHz VLSSr survey,][]{Intema2017, Rengelink1997, Lane2014}. The LOFAR Two-Metre Sky Survey \citep{Shimwell2017} is a new low-frequency survey aiming to map the entire northern sky. When completed, it will be two orders of magnitude deeper (in point-source sensitivity) and one order of magnitude higher in resolution than any current very large radio survey. The first 600 pointings (20\% of the total) of this survey have recently been observed, covering the 120--168~MHz band.\\

Here, we investigate the nature of the diffuse radio source in the galaxy cluster Abell 1931 using data from LOFAR (120-168 MHz), the Giant Metre Radio Telescope (GMRT) (325 MHz), the Very Large Array (VLA) (1--2 GHz) and the Chandra X-ray observatory (0.5--7~keV). We chose to study the galaxy cluster Abell~1931 because it was believed to have a relatively low mass (no ROSAT detection) and showed large-scale diffuse emission in LOFAR images that was not seen in other radio surveys. Steep and diffuse radio emission in low-mass clusters can yield new insights into the nature of the acceleration mechanism, the role of fossil electrons, and the origin of intracluster magnetic fields \cite{2017CQGra..34w4001V}.

In this paper we adopt a flat, $\Lambda$CDM cosmology with matter density $\Omega_M = 0.3$ and Hubble constant $H_0 = 67.8$ km s$^{-1}$ Mpc$^{-1}$ \citep{Planck2016}. Given the redshift of Abell 1931 ($z = 0.178$), the luminosity distance is $D_{\rm L} = 889$ Mpc and 1\arcsec corresponds to 3.109 kpc. All our images are in the J2000 coordinate system.

\begin{figure*}
 \includegraphics[width=\textwidth]{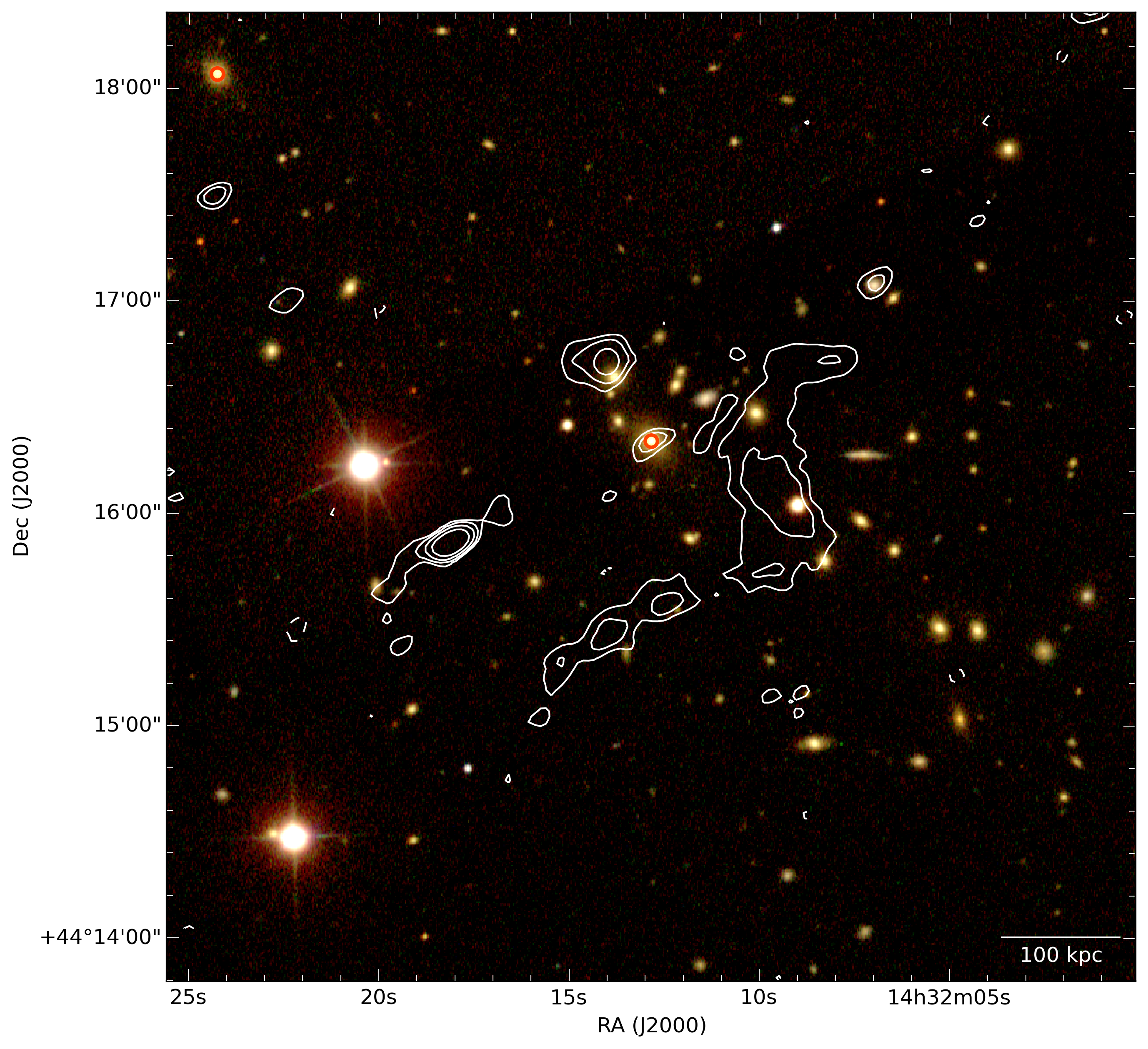}
 \caption{Overlay of SDSS $i,r,g$-band image with LOFAR HBA 120-168 MHz image of Abell 1931 with a noise of about 180 $\mu$Jy/beam. Contour levels are drawn at [-/+3 , 6 , 12, 24] $\sigma$ ($\sigma=180 \mu$Jy/beam). Galaxies with known redshifts are marked by red rings. The galaxy distribution seems to be elongated along a NE-SW axis, although many galaxies do not have confirmed redshifts.}
 \label{fig:LOFARirg}
\end{figure*}

\section{A new ultra-steep spectrum source in Abell~1931}
\subsection{LOFAR observations}

LOFAR is a digital, low-frequency radio interferometer that extends over large parts of Northern Europe and has its core in the Northern parts of the Netherlands \citep{vanHaarlem2013}. The  observation shown here was part of the LOFAR Two-meter Sky Survey (LoTSS; \citealt{Shimwell2017}) and was conducted with the high-band antennas (HBA) over a frequency range of 120--168 MHz. For a summary of all radio observations, see Tab.~\ref{tab:radio}.

After a complete direction-independent calibration of the data, we apply Facet Calibration. 
Facet Calibration is a direction-dependent calibration method for LOFAR in which the full field is tessellated into multiple facets \citep{2016ApJS..223....2V}. The full FOV that is approximately 3 degrees across is thus split into up to 50 facets. Each facet has its own calibrator source, chosen to have at least 0.3~Jy in flux density. Facet Calibration is applied using the software package FACTOR\footnote{\url{http://www.astron.nl/citt/facet-doc/}} which removes time and position-varying beam errors and ionospheric effects, which can be severe at these frequencies. It was ensured that the emission from the calibrator was deconvolved until the residual clean images were noise-like. For the pointing P218+45 (containing Abell 1931), the resulting image has  a noise level of $\sim$180 $\mu$Jy/beam and an angular resolution of about 5~\arcsec. The noise was determined from a circular region free of sources.\\

From the image, an extended, steep-spectrum source was found near the low-mass cluster Abell~1931. This source is faint and no radio emission is detected in the NVSS and VLSS surveys.   The source has a fairly irregular morphology with a largest linear size of about 550~kpc.\\

From the centre of the source (at approximately 14h32m11s,~+44d15.5m), there appear to be two arms: one extending to the South-East with a length of about 1\arcmin and a width of 0.25\arcmin  and one extending to the North with a length of about 1\arcmin and a width of 0.5\arcmin. There is no evidence for edge brightening or other distinct features such as jets or hot spots.\\

The diffuse emission appears to overlap with an overdensity of galaxies as seen in the Sloan Digital Sky Survey (SDSS). From the SDSS image, the source appears to be located in a very elongated cluster that is extended in the NE-SW direction, see Fig.~\ref{fig:LOFARirg}. The morphology suggests that this cluster is very likely undergoing a merger with the major axis of the source being perpendicular to the merger axis. 
The centre of the cluster is quoted at 14h31m59s,~+44d15.8m on NED\footnote{\url{https://ned.ipac.caltech.edu}}, but the SDSS galaxy distribution suggests that the cluster extends further to the East. A bright, red elliptical galaxy (SDSS J143212.84+441620.4), possibly the brightest cluster galaxy (BCG), has a radio counter-part and appears to be active. It has a spectroscopic redshift of $z=0.180$, an r-band magnitude of 16.94 and lies at coordinates 14h32m13s,~+44d16.26m.

The cluster redshift has been quoted as 0.170 \citep[photometric redshift only,][]{Lopes2004}. Three SDSS cluster galaxies (based on their colors) with spectroscopic redshift are available in the vicinity that yield a redshift of $z=0.178$, consistent with the photometric redshift. 

It is quite plausible that the BCG could be one of the suppliers of the relativistic electrons that produce the synchrotron emission in the diffuse source.  The connection of AGN with diffuse radio sources is the topic of a recent series of papers \citep[e.g.][]{Degasperin2014, Bonafede2014, Vanweeren2017, Degasperin2017, Wilber2017}.

From the primary-beam-corrected image, we measure a total flux density for the source within the 3-$\sigma$ contours at 143 MHz of 
$S_{143\,\mathrm{MHz}} = 39.7 \pm 8.0$ mJy (this is after correcting for the global flux offset relative to fluxes from the TIFR GMRT Sky Survey (TGSS) \citep{Intema2017}). The error includes a statistical error, estimated for the image noise in a nearby source-free region, and a flux-scale uncertainty of 20\%.

For a redshift of $z=0.178$, the luminosity distance is 888.9 Mpc, yielding a specific luminosity (including the k-correction as in \cite{Cassano2013}) of $P_{143\,{\rm MHz}} = 2.78  \pm 0.56 \times 10^{24}$ W~Hz$^{-1}$.

\subsection{GMRT observations}

Subsequently, we observed the cluster A1931 with the GMRT at 325 MHz under Director's discretionary time. It was observed on 15th June 06 2016 for 6 hours with an integration time of 8 seconds. The data were processed using the SPAM pipeline \citep[see][for details]{Intema2017}.

\begin{figure*}
 \includegraphics[width=0.33\textwidth]{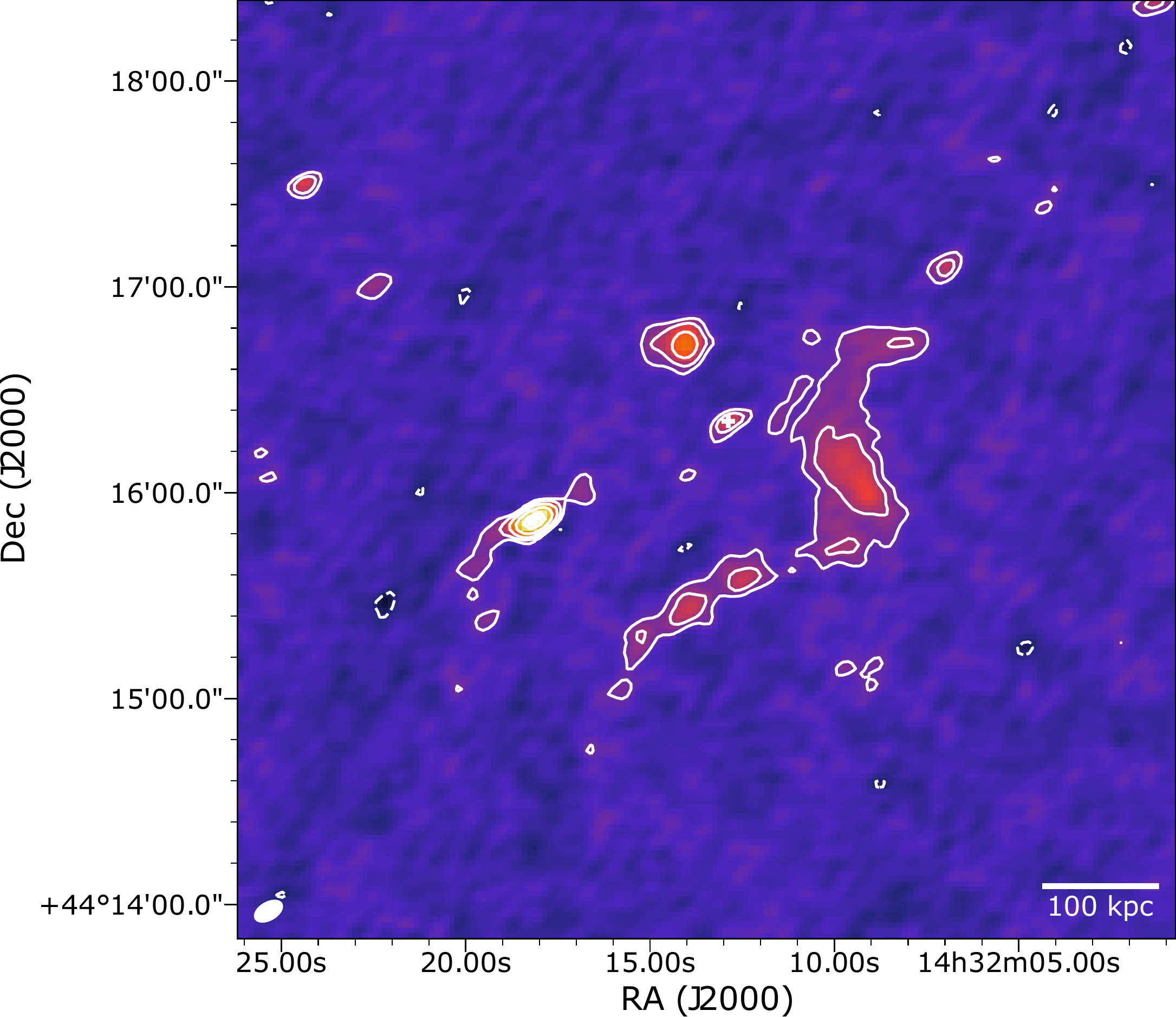}
 \includegraphics[width=0.33\textwidth]{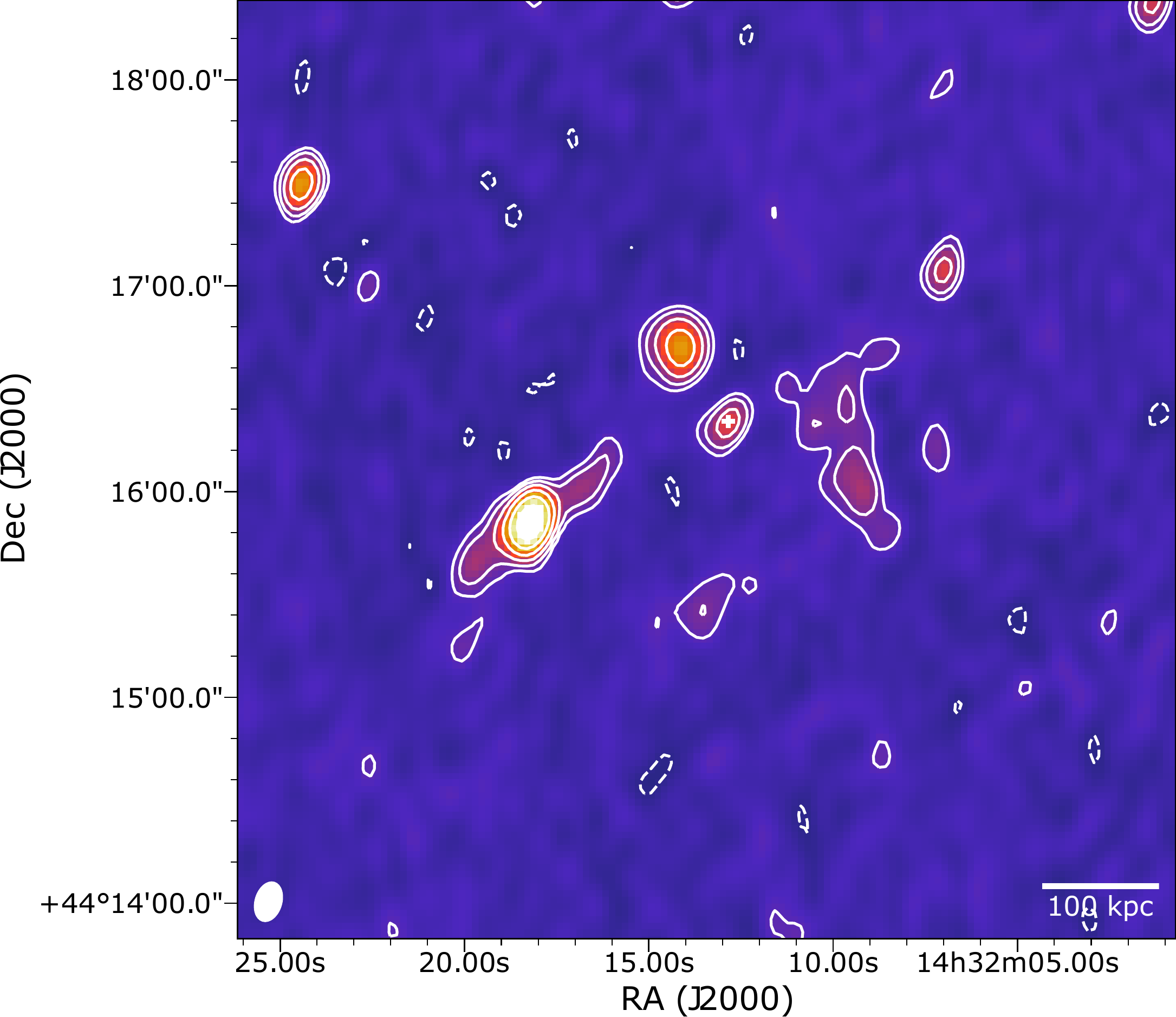}
 \includegraphics[width=0.33\textwidth]{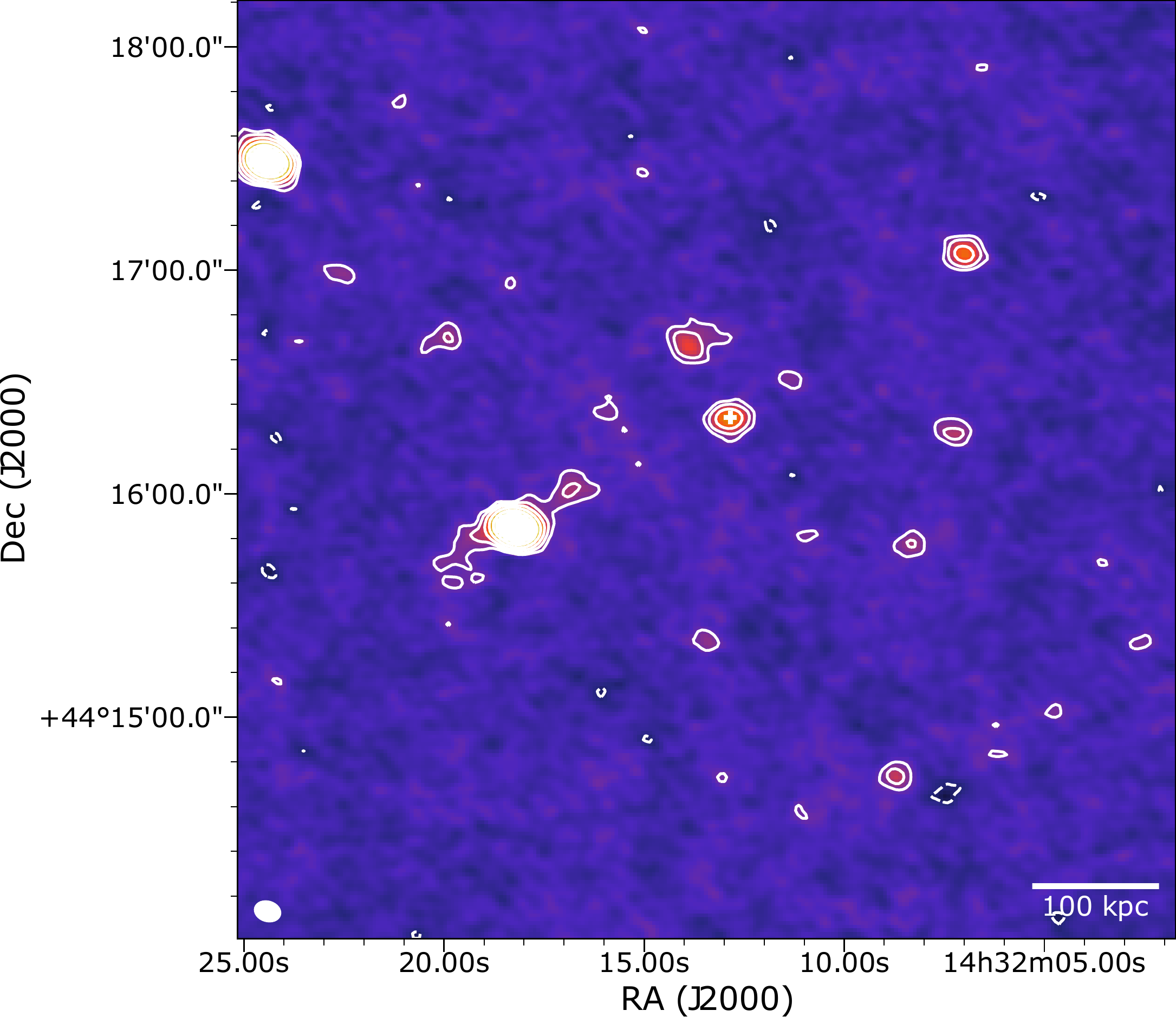}
 \caption{Left: LOFAR HBA 120-168 MHz image. Contour levels are drawn at [-/+3 , 6 , 12, 24] $\times\sigma$ ($\sigma=180$ $\mu$Jy/beam) and the restoring beam of $5.0\times 8.7$ \arcsec is shown by the ellipse in the lower-left corner. Middle: GMRT 325 MHz image. Contour levels are drawn at [-/+3 , 6 , 12, 24] $\times\sigma$ ($\sigma=55$ $\mu$Jy/beam); the restoring beam is $7.5\times 11.6$ \arcsec. Right: VLA 1-2 GHz B+C array image. Contour levels are drawn at [-/+3 , 6 , 12, 24] $\times\sigma$ ($\sigma=18$ $\mu$Jy/beam); the restoring beam is $5.2\times 7.0$ \arcsec. The cross marks the location of the BCG. No emission from the extended LOFAR source is detected in the VLA image.}
  \label{fig:GMRT}
\end{figure*}

Both, bandpass and absolute flux scale were calibrated using the source 3C286 as primary flux calibrator and applying the \cite{2012MNRAS.423L..30S} flux scale. We performed an initial phase-only calibration using a sky model generated from the surveys VLSSr, WENSS and the NVSS. SPAM uses AIPS tasks \citep[][]{2003ASSL..285..109G} to perform (faceted) wide-field imaging of the full field-of-view, followed by a series of self-calibration, wide-field imaging, and additional flagging of poor data where needed. Then we used the brightest sources in the primary beam to conduct a direction-dependent calibration.  Finally, SPAM fits ionospheric models to mitigate phase errors. The resulting image is shown in Fig.~\ref{fig:GMRT} 
and was made using slightly uniform weighting (ROBUST=-1). Note that the AIPS ROBUST parameter is a bit less strong than the CASA one.

We measure a total flux density for the diffuse emission from the GMRT image of $S_{325\,\mathrm{MHz}} = 3.85 \pm 0.82$ mJy (using the same region used for the LOFAR measurement, defined from the 3-$\sigma$ contours at 143 MHz), where we have adopted a flux scale uncertainty of 20\%. The corresponding power is $P_{325\,{\rm MHz}} = 2.7 \pm 0.58 \times 10^{23}$ W~Hz$^{-1}$.  The spectral index of the total source between 143 MHz and 325 MHz is therefore $\alpha_{143}^{325} = -2.86 \pm 0.36$. 

We note that, although the LOFAR and GMRT images are not matched precisely in terms of weighting and uv-coverage, both images should be equally sensitive to emission on the scales of the diffuse emission studied here ($\sim 2'$).

\subsection{VLA observations}

Abell 1931 was observed on May 29, 2017 and Dec 18 2017 with the Jansky Very Large Array (VLA) in C- and B-configurations, respectively. The observations were carried out in the 1--2~GHz L-band with a total on-source time of about 5 and 7~hrs, respectively.  We used the default frequency set-up which comprises two 512 MHz IF pairs (each comprising 8 contiguous subbands of 64 MHz) to cover the entire 1--2 GHz of the L-band receiver.
The data reduction was performed in CASA  \citep[v5.1;][]{2007ASPC..376..127M}, following the scheme described in \cite{2016ApJ...817...98V}. We briefly summarize the main steps: First, the data were Hanning smoothed and corrected for elevation dependent gains and antenna offset positions. Radio frequency interference (RFI) was removed with the 'tfcrop' mode in flagdata and in the final round with the {\tt AOFlagger} tool \citep{2010MNRAS.405..155O}. We calibrated the antenna delays, bandpass, cross-hand delays, and polarization leakage and angles using the primary calibrators 3C147 and 3C286. Gain solutions were then obtained for the secondary calibrator source and the relevant calibration tables were applied to the target field. Self-calibration was performed to refine the amplitude and phase calibration on the target.
All imaging was done with W-projection  \citep{2008ISTSP...2..647C,2005ASPC..347...86C} and MS-MFS \citep{2011A&A...532A..71R} with three Taylor terms ({\tt nterms=3}). For all self-calibration steps, we employed Briggs weighting \citep{Briggs1995}, with a {\tt robust} factor of 0.0. Clean masks were also made with the {\tt PyBDSF} package \citep{2015ascl.soft02007M}. After self-calibration, the data from the two array configurations were combined and a final gain calibration step was performed with a long solution interval (10~min) to align the datasets. The final images were corrected for the primary beam attenuation.

The resulting image is shown in Fig.~\ref{fig:GMRT}. For regular sources with spectral indices of  $\sim -0.5$, the VLA image goes significantly deeper than LOFAR the LOFAR image. Still, the diffuse LOFAR source remains undetected. We can place an upper limit on the value of the spectral index of the diffuse emission between 143 MHz and 1500 MHz by assuming the emission would fill the same region in the VLA image with a uniform surface brightness. Under these assumptions, any such emission must have a total flux of $S_{1-2{\rm~GHz}} < 0.52$~mJy to remain undetected in the VLA image, implying a spectral index between 143 MHz and 1.5 GHz of $\alpha_{143}^{1500} < -1.9$. With a redshift of $z=0.178$ this translates into an upper limit on the power of $P_{1.5{\rm~GHz}} < 4.9\times 10^{22}$ W~Hz$^{-1}$.

\begin{table*}
\caption{List of radio observations on A1931.}
\centering \tabcolsep 5pt
\begin{tabular}{l c c c c}
\hline\\
& LOFAR & GMRT & VLA B & VLA C \\\hline
Project code & LC3\_008 & ddtB221 & SI0264 & SI0264 \\
Observation date & Mar 16 2015 & June 6 2016 & Dec 18 2017 & May 29 2017\\
Central frequency & 143 MHz & 325 MHz & 1.5 GHz & 1.5 GHz\\
Bandwidth & 48 MHz & 33 MHz & 1 GHz & 1 GHz\\
On-source time & 8 h & 6 h & 7 h & 5 h\\
Noise & 180 $\mu$Jy/beam & 55 $\mu$Jy/beam & 18 $\mu$Jy/beam & 18 $\mu$Jy/beam  \\
Beam size & $5.0\times 8.7$ \arcsec & $7.5\times 11.6$ \arcsec & $5.2\times 7.0$ \arcsec & $5.2\times 7.0$ \arcsec
 \end{tabular}
\label{tab:radio}
\end{table*}

\section{X-ray observations of Abell~1931}

\begin{figure}
 \includegraphics[width=\columnwidth]{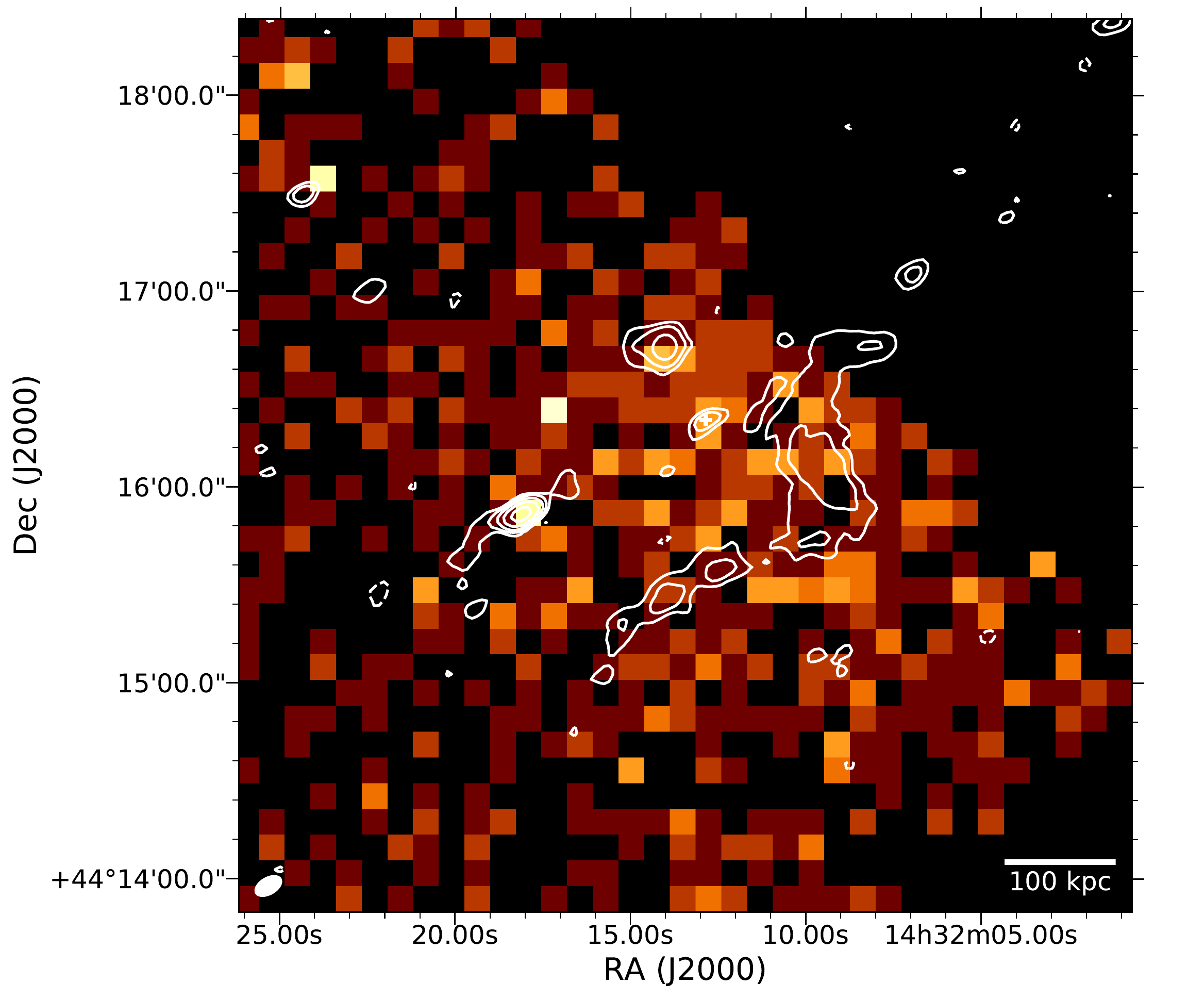}
 \caption{CHANDRA map with LOFAR contours overlaid. The LOFAR contours start at $3 \sigma$ (0.56 mJy/beam) and increase by a factor of 2. The image (0.5-2 keV, exposure-corrected) was binned such that each pixel is 16x16 original pixels in order to make the emission more visible. The cross marks the location of the BCG.}
 \label{fig:xraymap}
\end{figure}

After the discovery of the diffuse emission with LOFAR, we proposed for X-ray observations to determine the mass of the cluster. Abell~1931 was observed with the Chandra X-ray observatory (ObsId: 19569, VFAINT mode) in Cycle 18 on the 28th Sep. 2017 with a total exposure time of 40.5 ks, using the ACIS-I detector. The data were reprocessed with \textsc{CIAO} 4.9\footnote{See \url{cxc.harvard.edu/ciao/index.html}.} using \textsc{CALDB} 4.7.6\footnote{See \url{cxc.harvard.edu/caldb/index.html}.} and were corrected for known time-dependent gain and charge transfer inefficiency problems. Blank-sky background files were used for background subtraction.\footnote{See \url{http://asc.harvard.edu/contrib/maxim/acisbg/}.} The events files were filtered for flares using the \textsc{CIAO} script \emph{lc\_clean} to match the filtering used during the construction of the background files. A total of 5.7 ks was removed during filtering, resulting in a final exposure time of 34.8 ks. The background files were normalized to the count rate of the source image in the $10-12$ keV band (after filtering). The derived normalization ratio (in the sense of background/source) was 1.12. Lastly, point sources detected using the \textsc{CIAO} tool \emph{wavdetect} were removed. The X-ray map is shown in Fig.~\ref{fig:xraymap}. In accordance with the appearance of the galaxy distribution the ICM is elongated in the NE-SW direction.

For the spectral analysis, spectra were extracted using the \textsc{CIAO} script \emph{specextract}. For each spectrum, weighted responses were made, and a background spectrum was extracted in the same region of the CCD from the associated blank-sky background file. For the spectral fitting, \textsc{XSPEC} \citep{Arnaud1996} version 12.7.1 was used. X-ray spectra were extracted in a circular region centred on the approximate centroid of the cluster emission (at $\alpha =$ 14:32:14.7, $\delta =$ +44:16:22.2) with radius of 136 \arcsec containing $\approx 400$ background-subtracted counts. Gas temperatures were found by deprojecting these spectra using the PROJCT model in \textsc{XSPEC} with a single-temperature plasma component (MEKAL) absorbed by foreground absorption component (WABS), between the energies of 0.5 keV and 7.0 keV. In this fitting, the redshift was fixed to $z=0.178$, and the foreground hydrogen column density was fixed to $N_{\rm H} = 1.55 \times 10^{20}$~cm$^{-2}$, the weighted-average Galactic value from \cite{Dickey1990}. The metallicity of the MEKAL component was fixed to 0.4 solar, using \cite{Anders1989} abundances.

The total X-ray luminosity within a radius of 450 kpc is $L_X = (1.65 \pm 0.39) \times 10^{43}$ erg s$^{-1}$. Using the $M-L_{\rm X}$ relation by \cite{Reichert2011}, $M_{500}=1.39\times 10^{44} (L[10^{44}\, {\rm erg/s}])^{0.54}E(z)^{-0.93}$, this results in a mass of $M_{500}  = 5.25 \times 10^{13} M_{\odot}$.

The central temperature is $2.92_{-0.87}^{+1.89}$ keV. The $M-T$ relation, $M_{500}=0.29\times 10^{14} M_{\odot} (T[{\rm keV}])^{1.62}E(z)^{-1.04}$, then predicts a mass of $M_{500}  = 1.62_{-0.72}^{+2.04}\times 10^{14} M_{\odot}$. Unfortunately, the number of counts is not sufficient to assess the dynamical state of the cluster with the standard X-ray proxies as we cannot determine reliable centroid shifts or power ratios.\\

The cluster is not in the Planck cluster catalogue and is also not detected in the ROSAT All-Sky Survey (RASS). From the Planck non-detection, we conclude $M_{500} < 3 \times 10^{14}$~M$_{\odot}$ \citep{Planck2016}.

\section{Discussion and Modelling}

The diffuse radio source sits close to the centre of the cluster but the morphology of the source and the fact that we do not have sufficient data at 325 MHz to make a spectral index map does not allow an easy classification. The shape of the source makes it unlikely that it is a radio halo since it is neither round nor appears to fill the cluster centre. 

Also a gischt-type radio relic appears unlikely.  The steep spectrum is not consistent with a gischt relic that is associated with an active merger shock as those sources have integrated power-law spectra of $\sim -1$. Still, it is worth noting that the radio source appears to lie perpendicular to the merger axis of the cluster as suggested by the SDSS (see Fig.~\ref{fig:LOFARirg}) and also the X-ray image (Fig.~\ref{fig:xraymap}).

\begin{figure}
\includegraphics[width=\columnwidth]{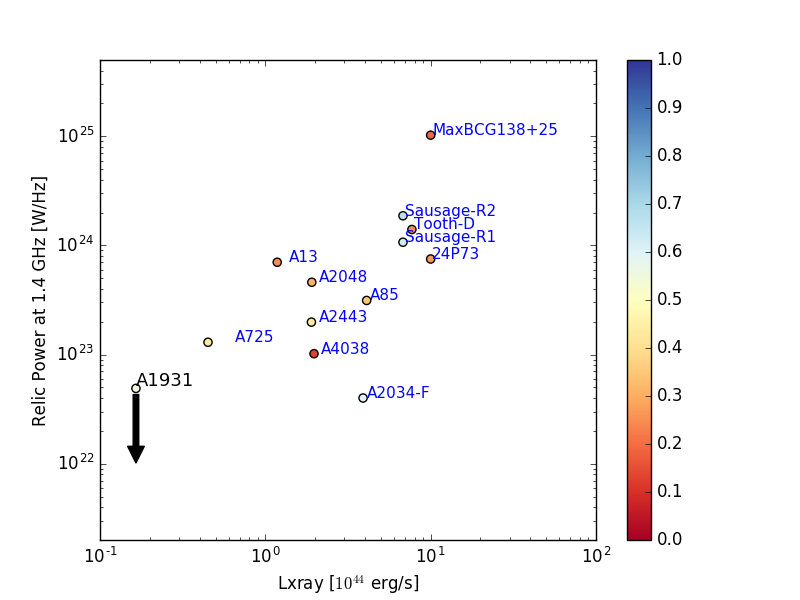}
\caption{Specific luminosity at 1.4 GHz of AGN/phoenix relics from Nuza et al. (2017) versus the X-ray luminosity of their host cluster in $10^{44}$ erg s$^{-1}$.  The color of the points represents their largest linear scale/size (LLS) in Mpc. The X-ray luminosity of the clusters 24P73 and MaxBCG 138+25 is not known.} 
\label{fig:fdg}
\end{figure}

So far, most radio relics have been found pre-dominantly in massive galaxy clusters.  In Fig.~\ref{fig:fdg}, we show the specific luminosity of AGN and phoenix-type radio relics at 1.4 GHz versus the X-ray luminosity (with the largest linear size color-coded). These sources are taken from the recent list compiled in \cite{Nuza2017} and this list is clearly not complete. The upper limit for Abell 1931 is shown and it lies at the lower end of both X-ray and radio luminosities. In many ways, this source is similar to the diffuse radio source in the merging galaxy cluster Abell 2443 (also shown in Fig.~\ref{fig:fdg}) which also has a steep spectrum, is offset from the cluster centre, has an irregular morphology, and is not clearly associated with any of the galaxies within the cluster \citep{Cohen2011}. This source is classified as a radio phoenix but the classification is uncertain.

Alternatively, the extended radio source in A1931 could be an AGN relic or remnant radio galaxy, similar to the one discovered by \cite{Brienza2017}. Even though there is no detectable connection in the LOFAR image, the likely BCG of the cluster could be the origin of the remnant radio emission. Bulk motions along the NE-SW directions could have moved the radio lobes from their source galaxy.\\

Customarily, the radio-synchrotron break frequency is used to estimate the age of the source. The break frequency $\nu_{\rm b}$ is related to the spectral age through
\begin{equation}
t = 1590 \frac{B^{1/2}}{B^2+B^2_{\rm CMB}} [\nu_{\rm b}(1 + z)]^{-1/2} \mathrm{Myr} ,
\end{equation}
where $B$ is given in $\mu$G and $\nu_{\rm b}$ in GHz. $B_{\rm CMB}$ is the magnetic field strength equivalent to the cosmic microwave background at the redshift of the source, $z$. Assuming the equipartition magnetic field of \cite{BK2005} with a proton-to-electron ratio of 100, we find $B_{\rm eq} = 1.9\,\mu$G for $\alpha =-2.86$. 
However, this calculation assumes an extremely steep power-law across the entire spectrum while in reality the spectrum may be curved. The injection spectrum at Lorentz factors where most of the energy density is contained is most likely to be flatter which would lead to a much weaker magnetic field. Also note that the associated magnetic pressure $B^2/8\pi\sim 1.4\times 10^{-13}$ erg cm$^{-3}$ is only a few percent of the central pressure derived from the X-ray data which is $p\sim 8 \times 10^{-12}$ erg cm$^{-3}$.

At this field strength, the radiative losses are dominated by inverse Compton losses. We do not know what the break frequency is but it must be smaller than $\sim 140$ MHz. Setting the break frequency to 140 MHz, we find that the spectral age must be larger than $\sim 220$ Myr.\\

Informed by hydrodynamical simulations of jet evolution, \cite{Hardcastle2018} have modelled the evolution of remnant radio galaxies. They found that when the supply of fresh particles ceases, radiative and adiabatic losses lead to an almost instantaneous disappearance of the high-frequency emission. 
In this semi-analytic model, it is assumed that a constant bi-polar jet expands into a spherically symmetric and isothermal cluster that follows the so-called universal pressure profile for clusters (Arnaud 1996). A constant fraction (assumed to be 0.4) of the energy supplied by the light jet goes into the internal energy of the lobe. The jets drive a shock into the ICM that follows the non-relativistic Rankine-Hugoniot relations. The magnetic field energy density is assumed to be a constant fraction (0.1) of the electron energy density at all times. The lobes behave as an adiabatic fluid with an adiabatic index, $\Gamma=4/3$ while the external medium has $\Gamma=5/3$. The electrons are injected with a constant power-law index (in energy) of 2.1 and the maximum and minimum Lorentz factor assumed for the injected electrons are 10 and $10^6$, respectively. For more details on the model, we refer the reader to \cite{Hardcastle2018}.\\

\begin{figure*}
\includegraphics[width=\columnwidth]{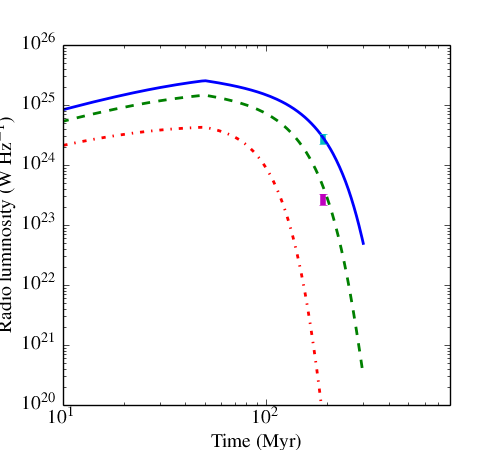}
\includegraphics[width=\columnwidth]{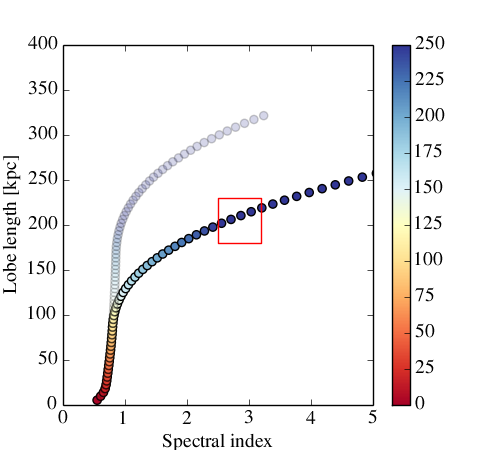}
\caption{Left: Luminosity at 143 MHz (solid blue), 325 MHz (dashed green) and 1.5 GHz (dotted red) as a function of time. Right: Length of lobe in kpc versus spectral index $\alpha_{143}^{325}$. Model assumes a universal pressure profile for a cluster of mass $M=10^{14} M_{\odot}$, a duration of the active jet of $t_{\rm off}=50$ Myr, a jet power of $Q=10^{36}$ W and a redshift of $z=0.178$. The errorbars and the red box, respectively, indicate measured values for A1931. Overlaid in transparent circles is an evolutionary path with all model parameters the same, only for $t_{\rm off}=200$ Myr.} 
\label{fig:remnant}
\end{figure*}

Varying the energy of the jet, $Q$, as well as the time when the jet is turned off, $t_{\rm off}$ as the only free parameters, we found that values of $t_{\rm off}=50$ Myr and $Q=10^{36}$ W gave a lobe length of $R=200$ kpc, a specific luminosity of $P_{143\,{\rm MHz}} = 1.26\times 10^{25}$ W/Hz and a spectral index between 143 MHz and 325 MHz of $\sim -2.8$ at a time of $\sim 190\pm 20$ Myr where the errors are determined from the right panel in Fig.~\ref{fig:remnant}. 
A jet power of $10^{36}$ W is expected for AGN host galaxies with stellar masses between $10^{11}M_{\odot}$ and $10^{12}M_{\odot}$\citep[e.g.]{Shabala2008}.
Fig.~\ref{fig:remnant} shows the radio luminosity as a function of time and a possible evolutionary path in the lobe length - spectral index $\alpha_{143}^{1500}$ space. For comparison, we have overlaid in transparent circles an evolutionary path for a model with $t_{\rm off}=200$ Myr, all other parameters staying the same. The width of the source with these model parameters is $\sim 50$ kpc which is also in line with the observations. This timescale is in reasonable agreement with the estimate based on the synchrotron break frequency alone (which ignores adiabatic effects etc.). Clearly, we have only presented one scenario within a somewhat idealised model and do not intend to perform an exhaustive modelling of this source given that the data is still very limited. The alleged centre of the diffuse radio source is located at a projected distance of about 100 kpc from the presumed BCG. Bulk flows in the cluster could have advected the remnant galaxy over a distance of 100 kpc within 120 Myrs assuming a velocity of 800 km/s. 

This rapid fading of the remnant sources has been invoked to explain the small number of remnant radio galaxies seen at low frequencies. Future searches are expected to reveal more about the effects and long-term evolution of radio jets in galaxy groups and clusters.

\section{Summary}

We have discovered a diffuse, steep spectrum radio source in the low-mass cluster Abell 1931. This source is only detected in LOFAR 120-168 MHz (HBA) and GMRT 325 MHZ images. From the primary-beam-corrected image, we measure a total flux density for the source at 143 MHz of $S_{143\,\mathrm{MHz}} = 39.7 \pm 8.0$ mJy  yielding a power of $P_{143\,{\rm MHz}} = 2.78 \pm 0.56 \times 10^{24}$ W/Hz. The source has a fairly irregular morphology with a largest linear size of about 550 kpc. The optical image suggests that the cluster is in the midst of a merger.

We measure a total flux density for the diffuse emission from the GMRT image of $S_{325\,\mathrm{MHz}} = 3.85 \pm 0.82$ mJy. The spectral index of the total emission between 143 MHz and 325 MHz is therefore $\alpha_{143}^{325} = -2.86 \pm 0.36$.

The diffuse radio source remained undetected in VLA L-band 1--2~GHz observations in C- and B-configuration. This leads to an upper limit on the flux density of $S_{1-2{\rm~GHz}} < 0.52$ mJy, implying a spectral index between 143 MHz and 1.5 GHz of $\alpha_{143}^{1400} < -1.9$. Unlike the sources in \cite{Brienza2016b} and \cite{Brienza2017} which are not in clusters, our source has a steep spectrum at low radio frequencies.\\

The cluster is not included in a Planck cluster catalogue and is also not detected in the ROSAT All Sky Survey. Dedicated Chandra X-ray observations of the cluster revealed a bolometric luminosity of $L_X = (1.65 \pm 0.39) \times 10^{43}$ erg s$^{-1}$ and a temperature of $2.92_{-0.87}^{+1.89}$ keV which implies a mass of $M_{\rm 500}\sim 10^{14} M_{\odot}$.\\

Based on size, spectral slope, power and morphology, we conclude that this source is likely to be a remnant radio galaxy. Modeling using a semi-analytic model by \cite{Hardcastle2018} suggest that the source has been produced by a source with a power of $Q=10^{36}$ W that was active for $t_{\rm off}=50$ Myr and is about 190 Myr old. 
A bright, red elliptical galaxy (SDSS J143212.84+441620.4), possibly the BCG of the cluster, has a radio counterpart indicating it is currently active AGN. Hence, it is conceivable that it is one of the sources of the relativistic electrons that produce the synchrotron emission.

Thus our observations afford a rare glimpse into faded sources whose role in feedback processes is not understood. Both, the length of time between outbursts as well as the duration of radio outbursts are central to quantifying the importance of AGN feedback. 

Finally, if this is indeed a merging galaxy cluster, it is interesting that there are no gischt-type relics or radio halos present.

\section*{acknowledgments}

MB acknowledges support from the grant BR2026/24 from the Deutsche Forschungsgemeinschaft (DFG).
AB acknowledges support from the ERC-StG 714245 DRANOEL. RJvW acknowledges support from the ERC Advanced Investigator programme NewClusters 321271 and the VIDI research programme with project number 639.042.729, which is financed by the Netherlands Organisation for Scientific Research (NWO).
LOFAR, the Low Frequency Array designed and constructed by ASTRON, has facilities in several countries, that are owned by various parties (each with their own funding sources), and that are collectively operated by the International LOFAR Telescope (ILT) foundation under a joint scientific policy.  The LOFAR software and dedicated reduction packages on https://github.com/apmechev/GRID\_LRT were deployed by the LOFAR e-infragroup, consisting of J.B.R. Oonk (ASTRON \& Leiden Observatory), A.P. Mechev (Leiden Observatory) and T. Shimwell (Leiden Observatory) with support from N. Danezi (SURFsara) and C. Schrijvers (SURFsara). This work has made use of the Dutch national e-infrastructure with the support of SURF Cooperative through grant e-infra160022. Support for this work was provided by the National Aeronautics and Space Administration through Chandra Award Number 19569 issued by the Chandra X-ray Center, which is operated by the Smithsonian Astrophysical Observatory for and on behalf of the National Aeronautics Space Administration under contract NAS8-03060. This research has made use of software provided by the Chandra X-ray Center (CXC) in the application packages CIAO, ChIPS, and Sherpa. We thank the staff of the GMRT who have made these observations  possible. GMRT is run by the National Centre for Radio Astrophysics of the Tata Institute of Fundamental Research. The NRAO is a facility of the National Science Foundation operated under cooperative agreement by Associated Universities, Inc.  We acknowledge helpful conversations with Raffaella Morganti and parts of code written by Martin Hardcastle published under https://github.com/mhardcastle.

\bibliographystyle{mnras}
\bibliography{A1931}

\bsp	
\label{lastpage}
\end{document}